\documentclass[fleqn,10pt]{wlscirep}

\title{Specific heat of 2D interacting Majorana fermions from holography}

\usepackage{xcolor}
\definecolor{tab20green}{rgb}{0.1725490196,0.6274509804,0.1725490196}
\definecolor{tab20red}{rgb}{0.8392156863,0.1529411765,0.1568627451}
\definecolor{tab20blue}{rgb}{0.1215686275,0.4666666667,0.7058823529}
\usepackage{amsfonts,amssymb,amsmath} 
\usepackage{graphicx}
\usepackage{dsfont}
\usepackage{soul}
\usepackage{ulem}
\usepackage{bm}

\graphicspath{{./figs/}}

\DeclareMathAlphabet\boldsymbolcal{OMS}{cmsy}{b}{n}

\newcommand{\tr}{\textrm{tr}}

\newcommand{\baleqn}{\begin{equation}\begin{aligned}[b]}
		\newcommand{\ealeqn}{\end{aligned}\end{equation}}
\newcommand{\baleqns}{\begin{equation*}\begin{aligned}}
		\newcommand{\ealeqns}{\end{aligned}\end{equation*}}

\newcommand{\be}{\begin{equation}}
	\newcommand{\ee}{\end{equation}}
\newcommand{\bq}{\begin{eqnarray}}
	\newcommand{\eq}{\end{eqnarray}}

\newcommand{\beq}{\begin{equation}}
	\newcommand{\eeq}{\end{equation}}
\newcommand{\beqa}{\begin{eqnarray}}
	\newcommand{\eeqa}{\end{eqnarray}}

\author[1]{Paolo Maraner}
\author[2]{Jiannis K. Pachos}
\author[3,*]{Giandomenico Palumbo}
\affil[1]{School of Economics and Management, Free University of Bozen-Bolzano, via Sernesi 1, 39100 Bolzano, Italy}
\affil[2]{School of Physics and Astronomy, University of Leeds, Leeds, LS2 9JT, United Kingdom}
\affil[3]{Center for Nonlinear Phenomena and Complex Systems,
	Universit\'e Libre de Bruxelles, CP 231, Campus Plaine, B-1050 Brussels, Belgium}

\affil[*]{giandomenico.palumbo@gmail.com}

	\begin{abstract}
	Majorana fermions are a fascinating medium for discovering new phases of matter. However, the standard analytical tools are very limited in probing the non-perturbative aspects of interacting Majoranas in more than one dimensions. Here, we employ the holographic correspondence to determine the specific heat of a two-dimensional interacting gapless Majorana system. 
	To perform our analysis we first describe the interactions in terms of a pseudo-scalar torsion field.
	We then allow fluctuations in the background curvature thus identifying our model with a $(2+1)$-dimensional Anti-de Sitter (AdS) geometry with torsion.
	By employing the AdS/CFT correspondence, we show that the interacting model is dual to a $(1+1)$-dimensional conformal field theory (CFT) with central charge that depends on the interaction coupling. This non-perturbative result enables us to determine the effect interactions have in the specific heat of the system at the zero temperature limit.
	
\end{abstract}
		
\begin{document}		
		
\flushbottom
\maketitle

\thispagestyle{empty}
	
\section*{Introduction}

Majorana fermions opened up a wealth of new possibilities in the behaviour of materials. As they support non-local quantum correlations they hold the promise of exhibiting novel quantum phases of matter. Moreover, they hold the key ingredient of quantum stability that is favourable for performing quantum computation~\cite{Kitaev,PachosBook}. Of particular interest are interacting Majorana fermions, which have diverse applications ranging from topological quantum technologies \cite{Fu}, to the modelling of black holes \cite{Sachdev,Kitaev2} and supersymmetric CFTs \cite{Grover, Giombi, Affleck}.

Relativistic behaviour of two-dimensional Majorana fermions in the presence of vortices that bound Majorana-zero-modes~\cite{Ivanov} has been recently identified in the low energy limit of various materials, such as A$_2$IrO$_3$ (A=Na, Li)~\cite{Singh} or $\alpha$-RuCl$_3$~\cite{Plumb,Banerjee,Winter}. As Majorana fermions are neutral we cannot probe charge response. But thermal response is still accessible. Recently, the heat capacity and thermal conductivities have been measured in various experiments~\cite{Hentrich,Kasahara} revealing the novel properties Majorana fermions induce to quantum matter. Due to the analytic tractability of free relativistic Majorana fermions they have enjoyed several rigorous numerical investigations of their thermodynamical properties~\cite{Nasu1,Nasu2,Nasu3,Self1,Self2}. 

Lately, an increasing amount of interest is focused on the effect interactions have in the topological characteristics of Majorana systems \cite{Kitaev3}. This is motivated by the high likelihood of having finite density of localised Majorana fermions in two-dimensional materials. The analytical study of such systems is formidable, while numerical analysis of interacting 2D Majorana systems faces numerous challenges. Most of the studies in two dimensions employ renormalisation group (RG) analysis \cite{Affleck2,Affleck3,Herbut} and mean-field theory~\cite{Affleck2,Franz,Franz2,Heath,Franz3}.

In this work, we focus on interacting Majorana fermions in two dimensions described by a Gross-Neveu-Yukawa model \cite{Zinn-Justin, Herbut2, Giombi, Affleck2,Xu,Herbut}. This model is defined by a spinless Majorana field coupled to a dynamical real pseudo-scalar field. 
To analyse it we first show that there exists a mapping between this interacting theory and an effective geometric theory with free fermions, where the underlying Riemann-Cartan geometry is entirely characterised by a non-null torsion. 
As we want to study the thermal effects of our microscopic model we then introduce a spacetime curvature in the system, encoded in the (torsionless) Ricci tensor.
The introduction of a non-trivial curvature in condensed-matter systems is physically justified by following the Luttinger's proposal \cite{Luttinger}.
In his derivation of the thermal transport coefficients in a near-equilibrium system, he showed that the effect of a thermal gradient is equivalent to that obtained from a fictitious gravitational potential. 
This idea has been implemented and extended in the case of topological systems supporting free Majorana and Dirac fermions in the recent years \cite{Stone,Ryu,Zhang}.
In these works Lorentz-invariant geometric models have been introduced in order to calculate quantum thermal effects, such as the thermal Hall effect on the boundary of the system. 
By following this approach we analyse the regime where anti-de Sitter geometry emerges in our model.
This choice is dictated by the very peculiar features of AdS, which allow us to employ the powerful dictionary of the holographic correspondence between gravitational models and quantum field theories. The extension of holography to different spaces is still an open problem in high-energy physics.
Importantly, the asymptotic boundary of the AdS solution in the higher-derivative geometric model that we have built, is equivalent to the asymptotic boundary of the AdS solution of a purely topological model.
Thus, by employing the AdS/CFT correspondence on this dual topological action, we obtain a $(1+1)$-dimensional CFT. Due to conformal anomaly the central charge of the CFT is proportional to the specific heat of the system when temperature, $T_{h}$, approaches zero~\cite{Affleck1986, Cardy1986, Cappelli}. 
Our approach then furnishes a powerful tool for analytically determining thermal signatures of interacting Majorana fermions. 
We find that the Gross-Neveu-Yukawa model of the 2D interacting Majoranas has a specific heat that grows linearly with temperature, a characteristic that is also met in interacting Dirac fermions. Moreover, we find that the specific heat directly depends on some couplings of the Majorana model while it is insensitive to others. These signatures can be directly used for the experimental verification of our results.

\section*{The 2D interacting Majorana model}

We start by considering interacting Majorana fermions in two dimensions at criticality. Such a phase can emerge, e.g. in a superconducting thin film placed in a perpendicular magnetic field on the top of a three-dimensional topological insulator~\cite{Affleck2} or at the boundary of a three-dimensional topological superconductor \cite{Grover}. The effective action in the low energy limit is given by
\begin{align}\label{eqn:action1}
S_{{\rm flat}}= \int  d^3 x\Big[
\frac{1}{2} \left(\bar \chi \gamma^\mu \partial_\mu \chi 
-\partial_\mu\bar \chi \gamma^\mu  \chi \right) + s \phi\bar\chi\chi
- \frac{1}{2} \partial_\mu\phi\partial^\mu\phi + \frac{r}{2} \phi^2  + \frac{u}{4} \phi^4\Big],
\end{align}
where $\mu=0,1,2$, indices are raised or lowered by $\eta_{\mu\nu}=\text{diag}(-1,1,1)$, the (real) Majorana representation $\gamma^\mu=(i \sigma^{y}, \sigma^{x},\sigma^{z})$, with $\sigma^{\mu}$ the Pauli matrices, $\chi$ is a two-component Majorana spinor, $\phi$ is a real pseudo scalar field and  $s$, $r$, $u$ are real parameters ($u>0$). 
This action is called the Gross-Neveu-Yukawa model that has a rich RG phase diagram, including an interacting supersymmetric CFT \cite{Grover,Giombi}. Upon integration of the pseudo-scalar field, one recovers a Gross-Neveu-like model that describes self-interacting Majorana fermions~\cite{Mihaila}. 

The first step to map this interacting model to a geometric one is to interpret the scalar field $\phi$ in terms of a torsion tensor. In Riemann-Cartan geometry, the torsion tensor $T_{\mu\nu\kappa}$ \cite{Hehl}, can be non-zero even in the flat space. In $2+1$ dimensions this tensor can be, in general, decomposed into a vector field, a traceless rank-2 symmetric tensor and a pseudo scalar field that we identify with $\phi$ \cite{Neto}.
In fact, as fermions couple only to the completely antisymmetric part of the torsion, we can choose, without loss of generality,
\be
T_{\mu\nu\kappa}=\frac{1}{3!}\,\epsilon_{\mu\nu\kappa}\phi, 
\label{eqn:torsion}
\ee
with $\epsilon_{\mu\nu\kappa}$ the Levi-Civita symbol.

\section*{Effective description with Anti-de Sitter geometry}

Even though the action~(\ref{eqn:action1}) provides an effective-field-theory description of interacting Majorana fermions, the study of its non-perturbative regime remains challenging. 
As Majorana fermions are neutral, they do not couple with a $U(1)$ gauge connection. Nevertheless, besides real pseudo-scalar fields, Majorana fermions can couple to a spacetime connection. 
The introduction of a non-trivial curvature is physically justified by following the Luttinger's proposal \cite{Luttinger} and its extensions in Majorana systems \cite{Stone,Ryu,Zhang}, in which thermal effects in condensed-matter systems
can be analyzed by introducing an effective curved spacetime.

Thus, we identify~(\ref{eqn:action1}) as the flat limit of a system of Majorana fermions in a Riemann-Cartan geometry. 
In this way, we can take advantage of the AdS/CFT correspondence \cite{Witten} that can provide non-perturbative results about strongly interacting systems \cite{Zaanen}. This formalism in the case of non-zero torsion \cite{Blagojevic,Blagojevic2,Klemm} has been successfully applied in the study of free massive Majorana fermions and thermal Hall effect on the gapped boundary of three-dimensional topological superconductors \cite{Palumbo-Pachos} (see also, \cite{Golan,Smith} for further applications of gravity and curved space to two-dimensional topological superconductors).

\subsection*{Curved background}

Let us consider the higher-derivative gravitational model in $2+1$ dimensions given by the action
\be
S_\text{RC} = \int d^3 x \sqrt{-g} \Big\{{1\over 2}\big[\bar \chi \gamma^\mu {\bm D}_\mu \chi - ({\bm D}_\mu\bar \chi)\gamma^\mu\chi\big]  + k_1{\bm R} + k_2{\bm R}^2 + k_3 {\bm R}^{\mu\nu} {\bm R}_{\mu\nu}\Big\},
\label{eqn:curved}
\ee
where the geometry is described by the metric compatible Riemann-Cartan connection 
\be 
{\boldsymbol{\Gamma}^\kappa}_{\mu\nu}={\Gamma^\kappa}_{\mu\nu}+
\frac{1}{2}\left({T_{\mu\nu}}^\kappa+{T_{\mu\,\,\,\nu}^{\,\,\kappa}}+
{T^\kappa}_{\mu\nu}\right),
\ee
with ${\Gamma^\kappa}_{\mu\nu}$ the Levi-Civita connection associated to the metric $g_{\mu\nu}$, ${T_{\mu\nu}}^\kappa={\boldsymbol{\Gamma}^\kappa}_{\mu\nu}-{\boldsymbol{\Gamma}^\kappa}_{\nu\mu}$ and $g= \det g_{\mu\nu}$.
Note that in curved space we have $\gamma^\mu = {e_a}^\mu\gamma^a$, where $\gamma^a$ are the gamma matrices defined as in (\ref{eqn:action1}) and with the dreibein ${e_a}^\mu$ fulfilling ${e_a}^\mu {e_b}^\nu g_{\mu\nu}=\eta_{ab}$.
The torsionful curvatures and spin connection are defined as 
${\boldsymbol{R}^\kappa}_{\lambda\mu\nu}=\partial_\mu{\boldsymbol{\Gamma}^\kappa}_{\nu\lambda}-\partial_\nu{\boldsymbol{\Gamma}^\kappa}_{\mu\lambda}+
{\boldsymbol{\Gamma}^\kappa}_{\mu\xi}{\boldsymbol{\Gamma}^\xi}_{\nu\lambda}-{\boldsymbol{\Gamma}^\kappa}_{\nu\xi}{\boldsymbol{\Gamma}^\xi}_{\mu\lambda}$, 
$\boldsymbol{R}_{\mu\nu}={\boldsymbol{R}^\kappa}_{\mu\kappa\nu}$, $\boldsymbol{R}=g^{\mu\nu}\boldsymbol{R}_{\mu\nu}$ and  
$\boldsymbol{ \Omega}_{\kappa ab}=\left(\partial_\kappa{e_a}^\mu+ {\boldsymbol{\Gamma}^\mu}_{\kappa\lambda}{e_a}^\lambda\right){e_b}^\nu g_{\mu\nu}$, respectively, so that the covariant derivative on spinors  is given by $\boldsymbol{D}_\mu = \partial_\mu + \boldsymbol{\omega}_\mu$, with $\boldsymbol{\omega}_\mu=\frac{1}{8}\boldsymbol{\Omega}_{\mu ab}[\gamma^a,\gamma^b]$.

Denoting by $R_{\mu\nu}$ and $R$ the Ricci and scalar curvatures associated to $g_{\mu\nu}$, we  obtain
\be
\boldsymbol{R}_{\mu\nu}=R_{\mu\nu}-\frac{1}{2}{\varepsilon_{\mu\nu}}^\kappa\partial_\kappa\phi+\frac{1}{2}\phi^2g_{\mu\nu},\hskip5pt
\boldsymbol{R}=R+\frac{3}{2}\phi^2.
\ee
As a result, in the flat limit, namely ${e_a}^\mu\to\delta_a^\mu$ that implies $g_{\mu\nu}\to \eta_{\mu\nu}$, the action $S_\text{RC}$ given in (\ref{eqn:curved}) reduces to the action $S_\text{flat}$ given in (\ref{eqn:action1}) when $k_1 = \frac{1}{3}r$,  $k_2 = \frac{1}{9}u-\frac{1}{3}$, $k_3=1$ and $s=\frac{3}{4}$. 
We have then shown that an interacting model living in a flat spacetime can be embedded in a more general covariant geometric model. In such a model the spacetime curvature is able to take into account quantum thermal effects that characterise the original microscopic model as we will show in more details in the next sections.

\subsection*{Equations of motion}

In order  to apply the AdS/CFT correspondence, we first need to determine whether the equations of motion 
\be
{\delta S_\text{RC}\over \delta\bar{\chi}}=0, \,\,{\delta S_\text{RC}\over\delta\phi}=0\,\,\text{and} \,\,{\delta S_\text{RC}\over\delta {e_a}^\mu}=0 ,
\ee
admit an AdS solution.
For this purpose we look for analytic solutions with constant curvature, $R\equiv K$, covariantly constant torsion so that $\phi\equiv T$, and maximally symmetric spinor fields $\chi$. This last condition is implemented by requiring the $\chi$ to satisfy the Killing equation \cite{Fried}
\be
\label{Keq}
D_\mu \chi=\lambda\gamma_\mu\chi,
\ee
with $D_\mu$ the torsionless covariant derivative on spinors and $\lambda$ a possibly complex number. In fact, the solutions of this equation, called Killing spinors, represent 
maximally symmetric solutions of the Dirac equation in a curved spacetime. Such solutions include the AdS geometries that we are interested in. 

As $\left[D_\mu,D_\nu\right]\chi=-\frac{K}{24}\left[\gamma_\mu,\gamma_\nu\right]\chi$ and (\ref{Keq}) implies $\left[D_\mu,D_\nu\right]\chi=\lambda^2\left[\gamma_\mu,\gamma_\nu\right]\chi$, we obtain that  $K=-24\lambda^2$. Therefore, real values of $\lambda$ imply a negative curvature, while purely imaginary ones a positive curvature. On the other hand, the equation of motion $\delta S_\text{RC}/\delta\bar{\chi}=0$ is satisfied by Killing spinors if and only if 
\be
\label{LambT}
\lambda=-\frac{1}{4}T, 
\ee
implying that $\lambda$ has to be real and the curvature negative
\be \label{K}
K=-\frac{3}{2}T^2.
\ee 
A direct consequence of the reality condition of  $\lambda$ is that $\bar{\chi}\chi\equiv M$ becomes constant in space as $\partial_\mu (\bar \chi \chi) = D_\mu (\bar \chi \chi) = (D_\mu\bar\chi)\chi + \bar\chi D_\mu\chi = (-\lambda^*+\lambda)\bar\chi\gamma_\mu\chi=0$.
By taking into account (\ref{Keq}), the remaining equations $\delta S_\text{RC}/\delta\phi=0$, $\delta S_\text{RC}/\delta {e_a}^\mu=0$ reduce to the algebraic equations in $T$ and $M$ 
\be
r\, T+\frac{3}{4}M = 0,\,\,\,\,\,\,
T\left(-\frac{1}{3}r\, T-\frac{1}{4}M\right) = 0.
\ee
The unique solution is given by 
\be
K = -\frac{27}{32}\left({M\over r}\right)^2,  \,\,\,\,\,\,  T=-\frac{3}{4}{M\over r},
\label{eqn:solutions}
\ee
corresponding to an AdS$_{2+1}$ with non-zero torsion for any $M\neq0$. We see that the Majorana ``condensation" parameter $M$ is related through (\ref{eqn:solutions}) to the radius of the AdS space which is a purely geometric quantity. This parameter, following Luttinger's approach, takes into account quantum thermal effects that we want to include in the model given in (\ref{eqn:action1}), as we shall see in the following. Moreover, the tight relation between Majorana fields and torsion exhibited by (\ref{eqn:solutions}) follows the paradigm of $3+1$ dimensions, where geometric torsion is generated by fermionic fields~\cite{Kibble}.

Differently from the higher-derivative gravity model without matter \cite{Blagojevic}, we have shown in our geometric model with matter that the torsion is not zero even at the level of the equations of motion. This is compatible with the general feature of gravitational models with fermion matter, where the spinor field is the source of torsion even at semi-classical level \cite{Hehl}.
Note that the AdS$_{2+1}$ space is defined in terms of groups by SO$(2,2)$, which contains the Lorentz group SO$(2,1)$.

\section*{Topological action}

The higher-derivative Riemann-Cartan action $S_\text{RC}$ (\ref{eqn:curved}) shares the same AdS solution with constant torsion with the following topological action \cite{Klemm}
\be
S_\text{topo} = \int d^3 x \sqrt{-g} \Big( \boldsymbol{R} + 2\Lambda + \alpha\, \varepsilon^{\mu\nu\kappa} T_{\mu\nu\kappa}\Big),
\label{eqn:topo2}
\ee
when $\Lambda = -\left({9 M\over 8 r}\right)^2$ and $\alpha={3\over 4}{M\over r}$. It is straightforward to show that the equations of motions of this action give us the AdS geometry with constant torsion and with $\Lambda<0$ the cosmological constant \cite{Klemm}. Moreover, both actions~(\ref{eqn:curved}) and~(\ref{eqn:topo2}) are time-reversal invariant due to the absence of the gravitational Chern-Simons term.
The action $S_\text{topo}$ is topological as it can be written in terms of Chern-Simons theories. In particular, the AdS$_{2+1}$ group decomposes as SO$(2,2)=\text{SO}(2,1)\times\text{SO}(2,1)$. So we can write
\be
S_\text{topo} = {t\over 8\pi} S_\text{CS}(A) - {\bar t\over 8\pi} S_\text{CS}(\bar A),
\label{eqn:topo}
\ee
where
\be
S_\text{CS}(A) = \int d^3 x\epsilon^{\mu\nu\kappa}\tr\Big(A_\mu \partial_\nu A_\kappa + {2 \over 3} A_\mu A_\nu A_\kappa\Big).
\ee
Here, the trace is taken on the Lorentz group SO$(2,1)$, the couplings are given by $t=\bar t = \pi\sqrt{2 \over 3}{16 \over 3 }{r \over M}$ and the two independent SO$(2,1)$ gauge connections $A$ and $\bar A$ are given by 
\be
A_\mu =\boldsymbol{\omega}_\mu +q e_\mu\,\,\text{and}\,\,\bar A_\mu =\boldsymbol{\omega}_\mu -\bar q e_\mu,
\ee 
where $e_\mu=\frac{1}{2}\gamma_{a}{e^{a}}_{\mu}$, $q = {3\over 8}{M\over r}\big(1+\sqrt{6}\big)$ and $\bar q =  {3\over 8}{M\over r}\big(1-\sqrt{6}\big)$. 
As action (\ref{eqn:topo2}) is topological but non-trivial, it implies that $A$ and $\bar A$ are pure gauge connections. This can be seen by calculating the equations of motion of~(\ref{eqn:topo}) with respect to $A$ and $\bar A$, respectively
\be
\epsilon^{\mu\nu\kappa}F_{\mu\nu}(A)=0, \hspace{0.3cm}\epsilon^{\mu\nu\kappa}F_{\mu\nu}(\bar A)=0,
\label{eqn:flat}
\ee 
where $F_{\mu\nu}(A)=\partial_{\mu}A_{\nu}-\partial_{\nu}A_{\mu}+[A_{\mu},A_{\nu}]$ is the curvature tensor of $A$ (the same holds for $\bar A$).
The AdS$_{2+1}$ solution with non-zero torsion corresponds to such pure-gauge configurations.

We explain now the reason why (\ref{eqn:curved}) and~(\ref{eqn:topo2}) share the same AdS sector even in the absence of any fermionic field in Eq.~(\ref{eqn:topo2}). First, we perform the variation of fermionic Lagrangian ${\cal L}_\text{fer}$ in Eq.(\ref{eqn:curved}) in terms of dreibeins ${e_a}^\mu$ obtaining
\be
\delta {\cal L}_\text{fer} = {G^a}_{\mu} \delta {e_a}^\mu,
\ee
with ${G^a}_{\mu}$ given by
\begin{align} 
\label{G}
{G^a}_{\mu}=\sqrt{-g}\Big[\frac{1}{4}\left(\bar{\chi}\gamma^{\nu}D_{\mu}\chi+\bar{\chi}\gamma_{\mu}D^{\nu}\chi\right){e^a}_{\nu}  
-\frac{1}{4}\left(D^{\nu}\bar{\chi}\gamma^{\nu}\chi+D^{\nu}\bar{\chi}\gamma_{\mu}\chi\right){e^a}_{\nu}-\nonumber\\
\frac{1}{2}\left(\bar{\chi}\gamma^{\kappa}D_{\kappa}\chi-D_{\kappa}\bar{\chi}\gamma^{\kappa}\chi\right){e^a}_{\mu}-\frac{3}{4}\phi \bar{\chi}\chi {e^a}_{\mu}-\nonumber\\
\frac{1}{4}{\varepsilon_{\mu}}^{\nu\kappa}\left(D_{\lambda}\bar{\chi}\gamma^{\lambda}-\frac{3}{4}\phi\bar{\chi}\right)\gamma_{\kappa}\chi {e^a}_{\nu}-\frac{1}{4}{\varepsilon_{\mu}}^{\nu\kappa}\bar{\chi}\gamma_{\kappa}\left(\gamma^{\lambda}D_{\lambda}\chi+\frac{3}{4}\phi\chi\right){e^a}_{\nu}
\Big].
\end{align}
Then we substitute the Killing condition (\ref{Keq}) and the other conditions (\ref{LambT}), (\ref{K}), (\ref{eqn:solutions}) that give rise to the AdS solution in Eq. (\ref{G}).
The resulting expression is given by
\be
\delta {\cal L}_\text{fer}\Big|_\text{AdS} \propto \sqrt{-g}\, \bar\chi \chi \lambda\, {e^a}_\mu \delta {e_a}^\mu = M \lambda \sqrt{-g}\, {e^a}_\mu \delta {e_a}^\mu .
\ee
Here, $M=\bar\chi\chi$ is constant as shown in the previous section.
Second, we note that variation of the cosmological constant term
\be
{\cal L}_{\Lambda} = \sqrt{-g}\, \Lambda,
\ee
with respect to the dreibeins is given by $\delta {\cal L}_{\Lambda} = -\Lambda \sqrt{-g}\, {e^a}_\mu \delta {e_a}^\mu$,
as $\delta  \sqrt{-g} = - \sqrt{-g} {e^a}_\mu \delta {e_a}^\mu$. 
Hence, the variation of the fermionic Lagrangian with the Killing condition and the AdS solution is equal to the variation of a cosmological constant term \be
\delta {\cal L}_\text{fer}\Big|_\text{AdS} = \delta {\cal L}_{\Lambda}
\ee 
for $\Lambda = -M\lambda$.
In other words, $\Lambda$ contains the information about Majorana fermions where now their dynamics is frozen due to the Killing  condition~(\ref{Keq}).
Importantly, the AdS solutions of both topological and higher-derivative actions share the same asymptotic boundary, which encodes the holographic CFT.
Moreover, any cosmological constant term does not contribute to the asymptotic boundary action or the corresponding CFT~\cite{Klemm} (see next section and Appendix).
Hence, to obtain the CFT that corresponds to $S_\text{RC}$ we can consider the AdS/CFT correspondence for $S_\text{topo}$ by following the treatment developed in Refs~\cite{Blagojevic2,Klemm} as shown next.

\vspace{0.3cm}

\section*{Holographic correspondence}

To apply the holographic correspondence to the topological action (\ref{eqn:topo}), we assume that the spacetime is diffeomorphic to $M_2\times {\cal R}$, a two-dimensional Minkowski spacetime and a radial part. We parametrise this space with the local coordinates $x^\mu=(x^i,\rho)$ ($i=0,1$), with $\rho$ the radial coordinate and with $M_2$ the spacetime on which the dual CFT lives.
We then solve the equations~(\ref{eqn:flat}) and expand the $(2+1)$-dimensional line element $ds^{2}$ by using the Fefferman-Graham expansion \cite{Klemm}
\be
ds^{2}=[e^{2\rho}g_{ij}+k_{ij}+k_{ji}+e^{-2\rho}\eta_{AB}{k^{A}}_{i}{k^{B}}_{j}]dx^{i}dx^{j}+l^{2}d\rho^{2},
\ee
where $A,B=0,1$ are the spacetime indices of the tangent space to the boundary $M_2$, $l = \sqrt{2\over 3}{4 \over 3}{r \over M}$, $k_{ij}$ is the extrinsic curvature on the asymptotic (1+1)-dimensional boundary and ${k^{A}}_{j}=e^{Ai}k_{ij}$, with $e_{Ai}$ the {\it zweibein}.
In the asymptotic limit $\rho\to \infty$, there appears a divergence similar to the standard ultraviolet divergences in quantum field theory.
This divergence can be regularised by adding a counter-term $S_\text{topo}^\text{c.t.}$, given by 
\be
S_\text{topo}^\text{c.t.}=-\frac{4}{l}\int_{M_{2}}\epsilon_{AB}\epsilon^{ij}{k^{A}}_{i}{e^{B}}_{j}.
\ee
As explained in Ref. \cite{Bala}, this subtraction depends only on the intrinsic geometry of the boundary. The counter-term is defined once and for all, differently from the
prescription involving embedding the boundary in a fixed reference frame.

In this way, ii the holographic energy-momentum tensor ${\tau_A}^i$ is given by
\be
{\tau_A}^i = {2\pi \over |e|}{\delta (S_\text{topo}+S_\text{topo}^\text{c.t.}) \over \delta {e^A}_i} = {2\pi \over |e|}{4\over l} \epsilon^{ij}\epsilon_{AB}{k^B}_j,
\ee
where $|e|$ is the determinant of ${e^A}_i$. Importantly, the boundary torsion is zero and the spin connection is then determined completely by the zweibein. By taking the trace of the energy-momentum tensor we find
\be
\tau = {e^A}_i{\tau_A}^i = 2\pi l R,
\ee
where $R$ is the 1+1-dimensional scalar curvature of the boundary. This represents the conformal anomaly of the dual CFT that can be rewritten as follows~\cite{Blagojevic2,Klemm}
\be
\tau={c\over 24} R,
\ee
with central charge
\be
c = 64\pi\sqrt{2\over 3}{r\over M}.
\label{eqn:c}
\ee
Hence, we have obtained that the central charge is proportional to the coupling $r$, which represents the effective mass of the pseudo-scalar field. Importantly, the central charge measures the degrees of freedom of the CFT and contains information about the corresponding specific heat when the temperature $T_{h}$ approaches zero.

\section*{Specific heat}

In two dimensions, free massless Dirac fermions in Dirac materials have specific heat, $C_\text{v}$, that scales quadratically with temperature, $T_\text{h}$ \cite{Balatsky}. This behaviour is also true for free massless Majoranas. In the presence of Coulomb interactions Dirac fermions have a specific heat that scales linearly with $T_\text{h}$. Hence, the change in the behaviour of the specific heat can witness the presence of interactions between Dirac fermions. On the other hand it is not known what the effect of interactions is on the specific heat of Majorana fermions. Moreover, as Majorana fermions are neutral their thermal behaviour is one of the few manifestations we have to probe their properties. Hence, it is of importance to understand which role interactions play in these lower-dimensional relativistic Majorana systems and how they change their thermal behaviour.

In the previous section we derived that the two-dimensional interacting gapless Majorana system given in (\ref{eqn:action1}) is effectively described by the dual CFT with central charge $c$ given by (\ref{eqn:c}). With this result at hand we are able to quantify the role interactions have in the specific heat of our model (\ref{eqn:action1}).
The specific heat $C_\text{v}$ of a CFT when $T_\text{h}\rightarrow 0$ is related to the conformal anomaly and its value is given by
\begin{eqnarray}
\label{heat}
C_\text{v}=\frac{\pi}{6}\, c\, K_\text{B}^{2} T_\text{h},
\end{eqnarray}
where $K_\text{B}$ is the Boltzmann constant (we have fixed the velocity of light to unity) \cite{Affleck1986, Cardy1986,Cappelli}.
This expression demonstrates that $C_\text{v}$ scales linearly with temperature, when $T_\text{h}\rightarrow 0$, reflecting the behaviour of interacting Dirac fermions. As our theoretical model given in (\ref{eqn:action1}) is defined in the continuum there is no energy scale to compare $T_\text{h}$. So the limit of small temperatures, where (\ref{heat}) is valid, is rather formal. An energy scale to compare $T_\text{h}$ can be obtained once we identify the underlying microscopic lattice model that gives effectively rise to the continuum model (\ref{eqn:action1}). Then the finite lattice spacing provides the energy scale for the temperature to be compared with~\cite{Calabrese}. 

Equation (\ref{heat}) represents the main result of our work that demonstrates the linear dependence of specific heat to temperature for the case of interacting Majorana particles \cite{Nasu3}. Our result paves the way for understanding the behaviour of two-dimensional interacting Majorana models in the non-perturbative regime. Our result applies only to the regime, $T_\text{h}\rightarrow 0$. So, at this stage we do not have a derivation of $C_\text{v}$ for our interacting model, $S_\text{flat}$, at finite temperature. However, the Luttinger's approach is very general and this implies that our results can be extended to the finite-temperature case by considering suitable geometric solutions. In fact, it is well known that the holographic correspondence allows to calculate finite-temperature thermal quantities in CFTs by considering black-hole solutions \cite{Zaanen}. In particular, our CFT at finite temperature would be dual to a BTZ (Banados-Teitelboim-Zanelli) black hole \cite{BTZ}, which is a solution of the topological action (\ref{eqn:topo2}) \cite{Blagojevic,Klemm}. Thus, we are confident that our approach can be extended to finite temperature. However, these calculations are beyond the scope of this work and the derivation of the finite-temperature specific heat for interacting Majoranas will be analysed in future work.

\section*{Conclusions}

Here, we considered a two-dimensional interacting Majorana system described by a Gross-Neveu-Yukawa model. This system can become of physical relevance. Experiments with dense configurations of Majoranas are currently realised that could give rise to interactions such as the ones described by  $S_\text{flat}$. The thermal effects in such systems are not readily amenable to current numerical or analytical methods. Nevertheless, by employing the holographic correspondence we were able to identify the specific heat of this system at the zero temperature limit. We found that the interactions change the behaviour of the specific heat. While the specific heat of free gapless Majorana systems scale quadratically with temperature, the presence of interactions makes this scaling linear. 

Being able to identify the change in the behaviour of specific heat  in a system from quadratic to linear could allow us to experimentally reveal the presence of strong interactions in novel topological phases supporting interacting Majorana fermions on their boundary \cite{Affleck2,Grover}. In fact, the existence of chiral Majorana fermions on the boundary of two-dimensional systems was recently demonstrated by measuring their thermal Hall conductivity \cite{Stern}. In the case of time-reversal invariant topological phases, the Hall conductivity is always zero and the specific heat is the main thermal observable related to topologically protected Majoranas.

Finally, a specific heat that scales linearly in temperature, $T_\text{h}$, was recently demonstrated for a two-dimensional system of free Majoranas that can support thermally excited vortices~\cite{Nasu3}. This behaviour is witnessed for temperatures above a characteristic temperature, where a crossover is induced to a regime where thermally excited vortices proliferate. This characteristic suggests that the linear behaviour of the specific heat is exclusively due to the Majorana binding vortices in the system. In contrast, we expect the linear dependence of $C_\text{v}$ in our system to occur at small temperatures, $T_\text{h}\to 0$. Our interacting Majorana system, rewritten as the Gross-Neveu-Yukawa model in (\ref{eqn:action1}), describes free Majorana fermions coupled to a dynamical pseudo scalar field $\phi$. In two dimensions we can view quantum vortices as point-like defects that can behave similarly to dislocations in lattice models. Static dislocations can be described by the Burgers vector, which becomes the torsion field in the continuum limit \cite{Vozmediano,Zaanen2,Fradkin}. It is intriguing to rigorously determine if an effective interpretation of the interacting Majorana system in terms of free Majoranas coupled to dynamical vortices is possible. Such a description would facilitate us to identify the mean field theory that faithfully describes the interacting model or to build efficient numerical approximations to simulate the 2D interacting system. We leave these tasks to a future project.

\section*{Appendix}
Here, we reproduce some of the results of Ref.~\cite{Klemm}, where they show that the cosmological constant does not contribute to the boundary stress-energy tensor.
We start considering the gravitational topological Lagrangian in Eq. (\ref{eqn:topo2}) rewritten in terms of form
\bq
{\cal L}_\text{topo}= e_{A}\wedge R^{A}+ \frac{\Lambda}{3}\epsilon^{ABC}e_{A}\wedge e_{B}\wedge e_{C}+\alpha e_{A}\wedge T^{A}= \nonumber \\
e_{A}\wedge d\omega^{A}+\epsilon^{ABC} e_{A}\wedge \omega_{B}\wedge \omega_{C}+ \frac{\Lambda}{3}\epsilon^{ABC}e_{A}\wedge e_{B}\wedge e_{C}+\nonumber \\
\alpha e_{A}\wedge de^{A}+\alpha \epsilon^{ABC}e_{A}\wedge \omega_{B}\wedge e_{C},
\eq
where $\wedge$ is the wedge product and $e_{A}$ and  $\omega_{A}$ are the dreibein and spin connection forms, respectively and $R^{A}$ and $T^{A}$ are the curvature and torsion forms, respectively.
We then have that
\begin{align}
\delta {\cal L}_\text{topo}=
\delta e_{A}\wedge d\omega^{A}+e_{A}\wedge d \delta \omega^{A}+ \epsilon^{ABC} \delta e_{A}\wedge \omega_{B}\wedge \omega_{C}+\nonumber \\
2\epsilon^{ABC} e_{A}\wedge \omega_{B}\wedge \delta \omega_{C}
+\Lambda\epsilon^{ABC}e_{A}\wedge e_{B}\wedge \delta e_{C}+
\alpha \delta e_{A}\wedge de^{A}+\alpha  e_{A}\wedge d \delta e^{A}+\nonumber \\
\alpha \epsilon^{ABC}e_{A}\wedge \delta\omega_{B}\wedge  e_{C}
+2\alpha \epsilon^{ABC}\delta e_{A}\wedge \omega_{B}\wedge e_{C}.
\end{align}
Only the terms in the above equation that can be written as total derivatives in the AdS spacetime $M_{3}$ contribute to the boundary action in $M_{2}=\partial M_{3}$.
In fact, we have that
\bq
\int_{M^{3}} e_{A}\wedge d \delta \omega^{A}+\alpha  e_{A}\wedge d \delta e^{A}=\nonumber \\
\int_{M^{3}} d(e_{A}\wedge \delta \omega^{A}+\alpha  e_{A}\wedge \delta e^{A})-\int_{M^{3}} (d e_{A}\wedge  \delta \omega^{A}+\alpha  d e_{A}\wedge \delta e^{A})= \nonumber \\
\int_{M^{2}} (e_{A}\wedge \delta \omega^{A}+\alpha  e_{A}\wedge \delta e^{A})-\int_{M^{3}} (d e_{A}\wedge  \delta \omega^{A}+\alpha  d e_{A}\wedge \delta e^{A}).
\eq
Thus, only the first two terms in the last line above contribute to the boundary action on $M_{2}$ because of the Stokes theorem (namely, they are total derivative).
Hence, the cosmological constant, which cannot be written as a total derivative term as well as the fermionic action with the Killing condition do not contribute to the boundary action. 
This implies that these terms do not contribute neither to the boundary stress-energy tensor.

\section*{Acknowledgments}

We thank Roman Jackiw and So-Young Pi for invaluable conversations and analytic derivations related to the geometrical part of the project as well as Agapitos Hatzinikitas and Matthew Horner for critical reading of the manuscript. This work was supported by the Royal Society International Exchanges grant IES/R2/170180 and the EPSRC grant EP/R020612/1. G. P. acknowledges ERC Starting Grant TopoCold for financial support.

\section*{Author contributions statement}

G. Palumbo and J. K. Pachos designed the research, P. Maraner, J. K. Pachos and G. Palumbo performed the calculations. All authors discussed the results and wrote the manuscript.

\section*{Additional information}

Competing financial and non-financial interests: The authors declare no competing financial and non-financial interests.

\end{document}